\documentstyle[12pt,fullpage,epsf]{article}
\begin{document}
\begin{center}
{\Large \bf Complete solution of Altarelli-Parisi evolution equation in
next-to-leading order and non-singlet structure function at low-x}\\
\vspace{.75cm}
{\large R.Rajkhowa  and
J.K.Sarma\footnote{E-mail:jks@tezu.ernet.in}}\\
\vspace{.25cm}
Physics Department, Tezpur University, Napaam,\\
Tezpur-784 028, Assam, India \\
\end{center}
\vspace{.75cm}

\begin{abstract}
We present complete solution of Altarelli-Parisi (AP) evolution 
equation in next-to-leading order (NLO) and obtain t-evolution of non-singlet 
structure function at low-x. Results are compared with HERA low-x and low-$Q^{2}$ 
data and also with those of complete solution in leading oder (LO) of AP evolution equation.
\end{abstract}
\vspace{.75cm}

\noindent {\large \bf 1. Introduction:}\\
\indent The Altarelli-Parisi (AP) evolution equations [1-4] are fundamental tools to study the 
t(=ln$(Q^{2}/\Lambda^{2})$)
and x-evolutions of structure functions, where x and $Q^{2}$ are Bjorken scaling variable and four momenta 
transfer in a deep inelastic scattering (DIS)  process [5] respectively and $\Lambda$ is the QCD cut-off
parameter. Though numerical solutions are available in the literature [6], the explorations of the possibility of 
obtaining analytical solutions of AP evolution equations are always interesting. In this connection,
complete solutions of AP evolution equations at low-x in leading order(LO) have  been obtained [7]. Its natural  
improvement will be the next-to-leading order (NLO) calculation.\\
\indent In this paper, we present complete solution of AP evolution equation in NLO at low-x and 
obtain t-evolution of non-singlet structure function. Results are compared with the HERA low-x low-$Q^{2}$
data, and also with those of complete solution in LO.
Here Section 1, Section 2 and Section 3 give the introduction, 
the necessary theory and the results and discussion respectively.\\
\vspace{.5cm}

\noindent{\large \bf 2. Theory:}\\
\indent The AP evolution equation for non-singlet structure function in NLO 
is [8]

$$\frac{\partial F_{2}^{NS}(x,t)}{\partial t}-
\frac{\alpha_{s}(t)}{2\pi}\left[\frac{2}{3}\{3+4ln(1-x)\}F_{2}^{NS}(x,t)-
\frac{4}{3}\int_{x}^{1}\frac{dw}{1-w}\{(1+w^{2})F_{2}^{NS}\left(\frac{x}{w},t\right)-
2F_{2}^{NS}(x,t)\}\right]$$
\begin{eqnarray}
-\left(\frac{\alpha_{s}(t)}{2\pi}\right)^{2}\left[(x-1)F_{2}^{NS}(x,t)\int_{0}^{1}f(w)dw+
\int_{x}^{1}f(w)F_{2}^{NS}\left(\frac{x}{w},t\right)dw\right]=0
\end{eqnarray}
where,
$$f(w)=\frac{16}{9}P_{F}(w)+2P_{G}(w)+\frac{2}{3}n_{f}P_{{N}_{F}}(w)+
\frac{2}{9}P_{A}(w).$$
The explicit forms of higher order kernels are [9]
$$P_{F}(w)=-\frac{2(1+w^{2})}{1-w}lnw\;ln(1-w)-\left(\frac{3}{1-w}+2w\right)lnw-\frac{1}{2}
(1+w)ln^{2}w-5(1-w),$$
$$P_{G}(w)=\frac{1+w^{2}}{1-w}\left(ln^{2}w+\frac{11}{3}lnw+\frac{67}{9}-
\frac{\pi^{2}}{3}\right)+2(1+w)lnw+\frac{40}{3}(1-w),$$
$$P_{{N}_{F}}(w)=\frac{2}{3}\left[\frac{1+w^{2}}{1-w}\left(-lnw-\frac{5}{3}\right)-2(1-w)\right]$$
and 
$$P_{A}(w)=\frac{2(1+w^{2})}{1+w}\int_{w/(1+w)}^{1/(1+w)}
\frac{dk}{k}ln\frac{1-k}{k}+2(1+w)lnw+4(1-w).$$
Running coupling constant in higher order has the form [10,11] 
$$\alpha_{s}(t)=\frac{4\pi}{\beta_{o}t}\left[1-\frac{\beta_{1}lnt}{\beta_{0}^{2}t}\right]$$
for one loop with 
$$\beta_{o}=\frac{33-2n_{f}}{3}\;{\rm{and}}\;\beta_{1}=\frac{306-38n_{f}}{3},$$
$n_{f}$ being the number of flavours.\\

\indent Using Taylor expansion method [12] and neglecting higher order terms 
as discussed in our earlier works [7,13,14], $F_{2}^{NS}(x/w,t)$ 
can be approximated for low-x as
\begin{eqnarray}
F_{2}^{NS}\left(\frac{x}{w},t\right)\simeq F_{2}^{NS}(x,t)+x\sum_{l=1}^{\infty}u^{l}\frac{\partial F_{2}^{NS}(x,t)}{\partial x}
\end{eqnarray}
where 
$$x=1-w\;{\rm{and}}\;\frac{x}{1-u}=x\sum_{l=0}^{\infty}u^{l}.$$

\indent Putting equation (2) in equation (1) and performing u-integrations we get,

$$\frac{\partial F_{2}^{NS}(x,t)}{\partial t}-\left[\frac{\alpha_{s}(t)}{2\pi}
A(x)+\left(\frac{\alpha_{s}(t)}{2\pi}\right)^{2}B(x)\right]\frac{\partial F_{2}^{NS}(x,t)}{\partial x}
$$
\begin{eqnarray}
-\left[\frac{\alpha_{s}(t)}{2\pi}C(x)+\left(\frac{\alpha_{s}(t)}{2\pi}\right)^{2}D(x)\right]F_{2}^{NS}(x,t)=0
\end{eqnarray}
where,
$$A(x)=\frac{2}{3}\{-2xlnx+x(1-x^{2})\},$$
$$B(x)=x\int_{x}^{1}\frac{1-w}{w}f(w)dw,$$
$$C(x)=\frac{2}{3}\{3+4ln(1-x)+(x-1)(x+3)\}$$
and
$$D(x)=-\int_{0}^{x}f(w)dw+x\int_{0}^{1}f(w)dw.$$
For a possible solution, we assume that
$$\left(\frac{\alpha_{s}(t)}{2\pi}\right)^{2}= k\left(\frac{\alpha_{s}(t)}{2\pi}\right)$$
where, k is a numerical parameter  to be obtained from the particular $Q^{2}$-
range under study. By a suitable choice of k we can reduce the error to a 
minimum. Now equation (3) can be recast as
\begin{eqnarray}
\frac{\partial F_{2}^{NS}(x,t)}{\partial t}-
P(x,t)\frac{\partial F_{2}^{NS}(x,t)}{\partial x}-Q(x,t)F_{2}^{NS}(x,t)=0,
\end{eqnarray}
where,
$$P(x,t)=\frac{\alpha_{s}(t)}{2\pi}[A(x)+kB(x)]$$
and
$$Q(x,t)=\frac{\alpha_{s}(t)}{2\pi}[C(x)+kD(x)].$$
The general solution of equation (4) is
$$F(U,V)=0$$
where, F is an arbitrary function and 
$$U(x,t,F_{2}^{NS})=C_{1}\;{\rm {and}}\;V(x,t,F_{2}^{NS})=C_{2},$$
where $C_{1}$ and $C_{2}$ are constants, form a solution of equations
\begin{eqnarray}
\frac{dx}{P(x,t)}=\frac{dt}{-1}=\frac{dF_{2}^{NS}(x,t)}{-Q(x,t)}.
\end{eqnarray}
Solving equation (5) we obtain,
$$U(x,t,F_{2}^{NS})=t^{(b/t+1)}\;exp\left[\frac{b}{t}+\frac{N(x)}{a}\right]$$
and
$$V(x,t,F_{2}^{NS})=F_{2}^{NS}(x,t)\;exp[M(x)]$$
where 
$$a=\frac{2}{\beta_{o}},\;b=\frac{\beta_{1}}{\beta_{o}^{2}},$$
$$N(x)= \int\frac{dx}{A(x)+kB(x)}$$
and
$$M(x)=\int\frac{C(x)+kD(x)}{A(x)+kB(x)}dx.$$

\indent If U and V are two independent solutions of equation (4) and if $\alpha$  
and$\beta$ are arbitrary constants, then  
$$V=\alpha U+\beta$$ 
is called a complete solution of equation (4). Then the complete solution [12]
$$F_{2}^{NS}(x,t)\; exp[M(x)]=\alpha\left[t^{(b/t+1)}\;exp\left(\frac{b}{t}+\frac{N(x)}{a}\right)\right]+\beta$$
is a two-parameter family of planes. The one parameter family 
determined by taking $\beta=\alpha^{2}$ has equation 
\begin{eqnarray}
F_{2}^{NS}(x,t)\; exp[M(x)]=\alpha\left[t^{(b/t+1)}\;exp\left(\frac{b}{t}+\frac{N(x)}{a}\right)\right]+\alpha^{2}.
\end{eqnarray}
Differentiating equation (6) with respect to $\alpha$, we get
$$\alpha =-\frac{1}{2}t^{(b/t+1)}\;exp\left[\frac{b}{t}+\frac{N(x)}{a}\right].$$ 
Putting the value of $\alpha$ again in equation (6), we obtain
$$F_{2}^{NS}(x,t)\;exp[M(x)]=-\frac{1}{4}\left[t^{(b/t+1)}\;exp\left(\frac{b}{t}+\frac{N(x)}{a}\right)\right]^{2}.$$ 
Therefore,
\begin{eqnarray}
F_{2}^{NS}(x,t)=-\frac{1}{4}t^{2(b/t+1)}\;exp\left[\frac{2b}{t}+\frac{2N(x)}{a}-M(x)\right]. 
\end{eqnarray}
Now, defining
$$F_{2}^{NS}(x,t_{o})=-\frac{1}{4}t_{o}^{2(b/t_{o}+1)}\;exp\left[\frac{2b}{t_{o}}+\frac{2N(x)}{a}-M(x)\right].$$ 
at $t=t_{o}$, where $t_{o}=ln(Q_{o}^{2}/\Lambda^{2})$ at any lower value $Q=Q_{o}$, we get from equation (7)
\begin{eqnarray}
F_{2}^{NS}(x,t)=F_{2}^{NS}(x,t_{o})\left(\frac{t^{(b/t+1)}}{t_{o}^{(b/t_{o}+1)}}\right)^{2}\; exp\left[2b\left(\frac{1}{t}-\frac{1}{t_{o}}\right)\right]
\end{eqnarray}
which gives the t-evolution of non-singlet structure function 
$F_{2}^{NS}(x,t)$ in NLO.
\indent In an earlier communication [7], we suggested that for low-x in LO 
\begin{eqnarray}
F_{2}^{NS}(x,t)=F_{2}^{NS}(x,t_{o})\left(\frac{t}{t_{o}}\right)^{2}.
\end{eqnarray}
We observe that if b tends to zero, then equation (8) tends to equation(9),
i.e., solution of NLO equation goes to that of LO equation. Physically b
tends to zero means number of flavours is high.  \\ 
\indent Again defining,
$$F_{2}^{NS}(x_{o},t)=-\frac{1}{4}t^{(b/t+1)}\;exp\left[\frac{2b}{t}+\frac{2N(x)}{a}-M(x)\right]_{x=x_{o}},$$
we obtain from equation (7)
\begin{eqnarray}
F_{2}^{NS}(x,t)=F_{2}^{NS}(x_{o},t)\;exp\int_{x_{o}}^{x}\left[\frac{2}{a}
.\frac{1}{A(x)+kB(x)}-\frac{C(x)+kD(x)}{A(x)+kB(x)}\right]dx
\end{eqnarray}
which gives the x-evolution of non-singlet structure function $F_{2}^{NS}(x,t)$ in NLO.\\  
\indent Proton and neutron structure functions measured in deep inelastic 
electro-production can be written in terms of singlet and non-singlet quark
distribution functions  as
$$F_{2}^{p}(x,t)=\frac{5}{18}F_{2}^{S}(x,t)+\frac{3}{18}F_{2}^{NS}(x,t)$$
and
$$F_{2}^{n}(x,t)=\frac{5}{18}F_{2}^{S}(x,t)-\frac{3}{18}F_{2}^{NS}(x,t).$$
These equations give
$$F_{2}^{NS}= 3(F_{2}^{p}-F_{2}^{n})$$
from which we can calculate experimental values of $F_{2}^{NS}$ in t and x 
ranges given in $F_{2}^{p}$ and $F_{2}^{n}$.\\ 
\vspace{.5cm}

\noindent{\large \bf 3. Results and Discussion:}\\
\indent In the present paper, we compare our results of t-evolution of 
non-singlet structure functions from equation (8) with the HERA low-x and
low-$Q^{2}$ data [15]. Here proton and neutron structure functions are
measured in the range $2\leq Q^{2}\leq 50\;GeV^{2}$.
Moreover here $P_{T}\leq 200\;MeV$, where $P_{T}$
is the transverse momentum of the final state baryon.
We consider number of flavours $n_{f}$=4.\\ 

\indent In figures 1(a-c) we present our results of t-evolution of 
non-singlet structure functions $F_{2}^{NS}$ (solid lines) for the 
representative values of x given to test the evolution equation (8) in next-to-leading order.
Agreement is found to be good. 
In the same figures we also plot the results of t-evolutions of non-singlet 
structure functions $F_{2}^{NS}$
(dashed lines) for the complete solutions from equation (9) 
in leading order. Data points at lowest-$Q^{2}$ values in the figures are 
taken as inputs. 
We observe that t-evolutions are slightly steeper in NLO calculations 
then those of LO. We can also calculate x-evolution of non-singlet
structure functions at low-x from equation (10). But it involves complicated
triple integrations and we keep it as our subsequent work.   \\

\indent In figure 2 we plot ${T(t)}^{2}$ and kT(t) against $Q^{2}$ in the
$Q^{2}$ range $2\leq Q^{2}\leq 50\;GeV^{2}$ as required by our data used[15].
Though the explicit value of k is not necessary in calculating t-evolution
of $F_{2}^{NS}$, yet we observe that for k= 0.027, errors become minimum and
varies from 17.68\% to -13.68\% in the $Q^{2}$ range
$2\leq Q^{2}\leq 50\;GeV^{2}$.\\  
\vspace{.5cm}

\noindent{\large \bf Acknowledgement:}\\
\indent We are grateful to G. A. Ahmed of Department of Physics and N. Basumatary
of Department of Information Technology of Tezpur University for the help
in the computational part of this work. One of us (JKS) is grateful to UGC,
New Delhi for the financial assistance to this work in the form of a major research project.\\ 
\vspace{.5cm}

\noindent {\large \bf References:}\\
$[1]$G. Altarelli and G. Parisi, Nucl. Phys. B 126 (1977) 298.\\
$[2]$G. Altarelli, Phy. Rep. 81 (1981) 1.\\
$[3]$V.N. Gribov and L.N. Lipatov, Sov. J. Nucl. Phys. 20 (1975) 94.\\
$[4]$Y.L. Dokshitzer, Sov. Phy. JETP 46 (1977) 641.\\
$[5]$F. Halzen and  A. D. Martin, Quarks and Leptons: An Introductory
   Course in Modern Particle Physics, John and Wiley (1984).\\
$[6]$M. Miyama and S. Kumano, hep-ph/9508246 (1995).\\
$[7]$R. Rajkhowa and J.K. Sarma, hep-ph/0202263 (2002).\\
$[8]$A. Deshamukhya and D.K. Choudhury, Proc. 2nd Regional Conf. Phys. Research in North-East,
Guwahati, India, October, 2001 (2001) 34.\\
$[9]$G.Curci, W.Furmanski and R.Petronzio, Nucl. Phys. B175 (1980) 27.
\\
$[10]$L.F. Abbott and R.M. Barnett, Ann.Phys. 125 (1980) 276.\\
$[11]$A. Saikia and D.K. Choudhury, Pramana-J. Phys. 38 (1992) 313.\\
$[12]$F. Ayres Jr., Differential Equations, Schaum's Outline Series, McGraw-Hill (1952).\\
$[13]$J.K. Sarma and B. Das, Phy. Lett. B304 (1993) 323.\\
$[14]$J.K. Sarma, D.K. Choudhury and G.K. Medhi, Phys. Lett. B403 (1997) 139.\\
$[15]$M. Arneodo et al., hep/961031, NMC, Nucl. Phy. B483 (1997).\\

\begin{figure}[htbp]
\begin{center}
\leavevmode
\centerline{\epsfbox{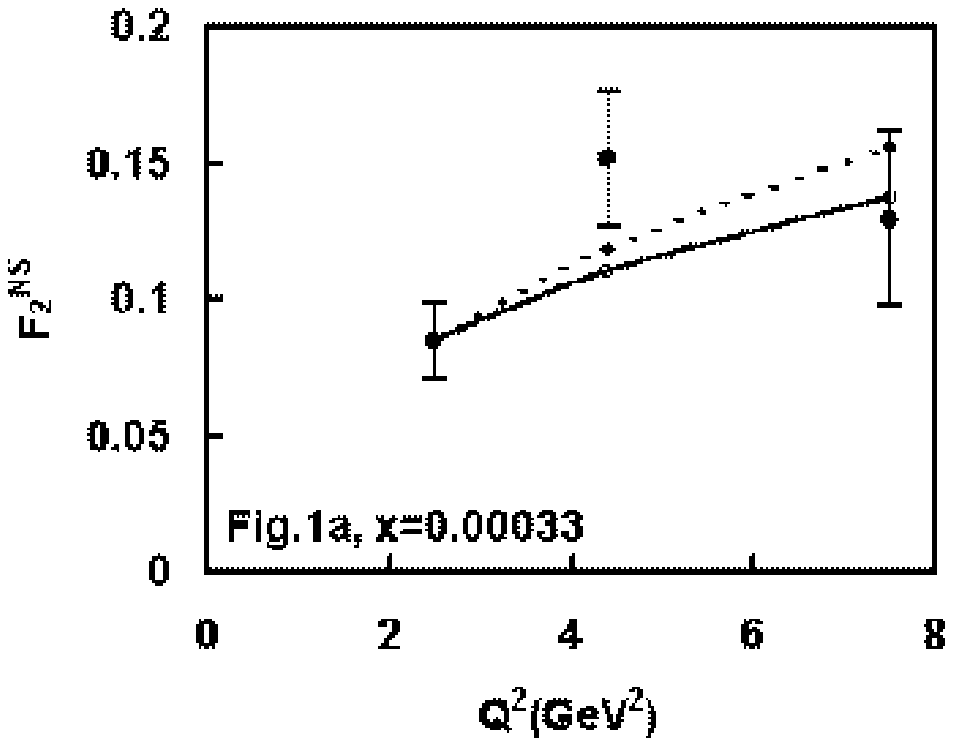}}
\end{center}
\end{figure}

\begin{figure}[htbp]
\begin{center}
\leavevmode
\centerline{\epsfbox{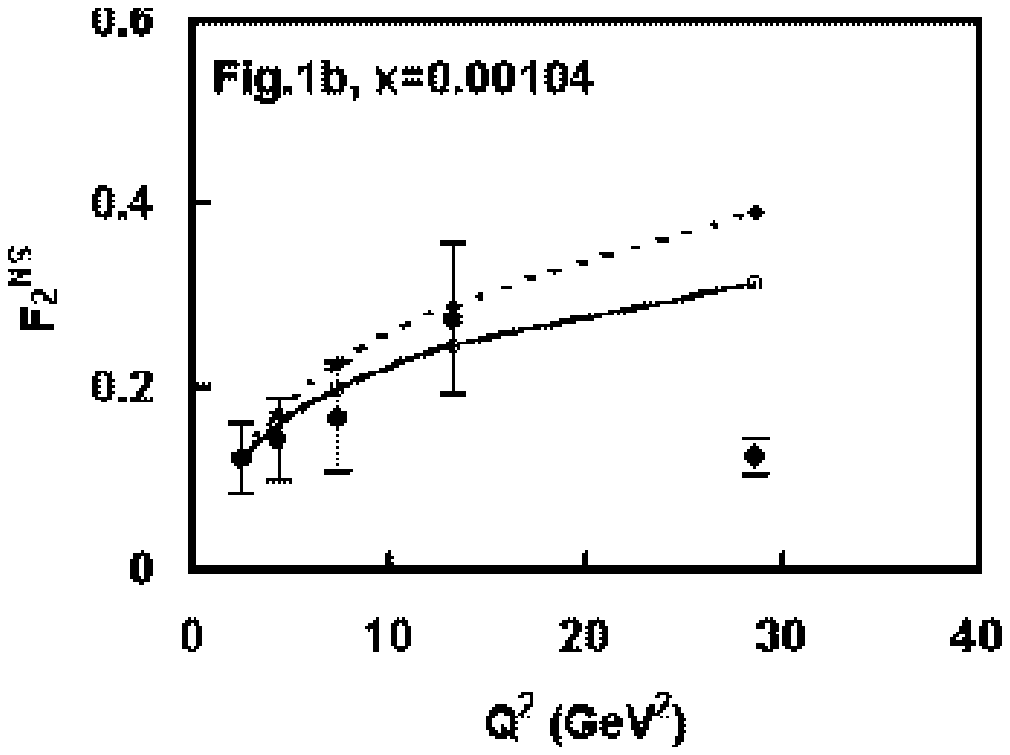}}
\end{center}
\end{figure}

\begin{figure}[htbp]
\begin{center}
\leavevmode
\centerline{\epsfbox{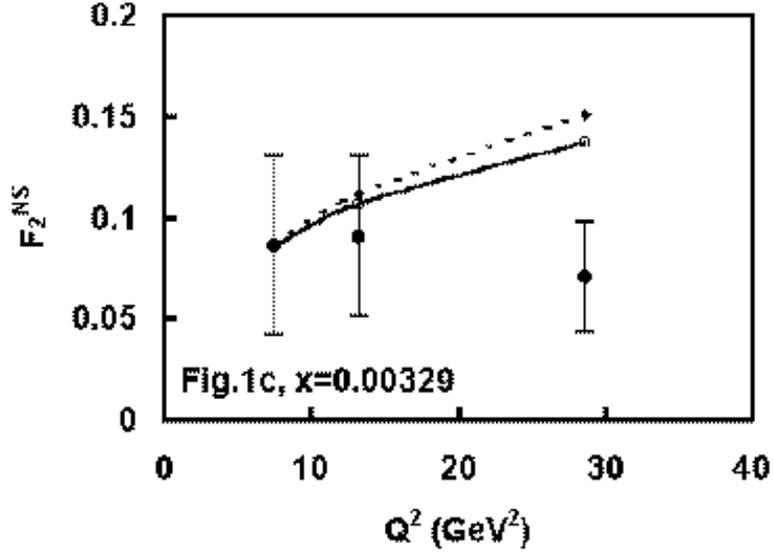}}
\end{center}
\caption {t-evolutions of non-singlet structure functions 
$F_{2}^{NS}(x,t)$(solid lines) for the representative values of x given in 
the figures. Data points at lowest-$Q^{2}$ values in the figures are taken as
input to test NLO t-evoluation of non-singlet structure functions 
$F_{2}^{NS}$ from  equation (8). In the same figures we also plot the results
of t-evolutions of non-singlet structure functions $F_{2}^{NS}$ 
(dashed lines) for LO from equation (9).}
\end{figure}

\begin{figure}[htbp]
\begin{center}
\leavevmode
\centerline{\epsfbox{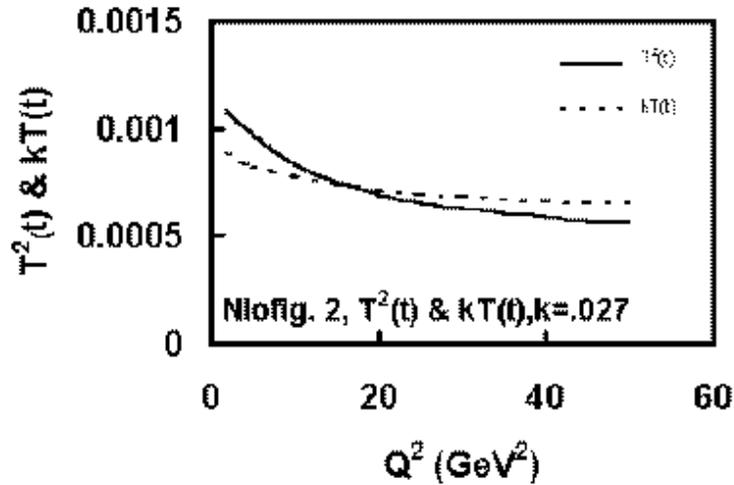}}
\end{center}
\caption {${T(t)}^{2}$ and kT(t) against
$Q^{2}$ in the range $2\leq Q^{2}\leq 50\;GeV^{2}$ for k=0.027.}
\end{figure}
\end{document}